\documentclass{research4cacm}

\renewcommand{\footnotesize}{\scriptsize}
\usepackage{color}
 \usepackage{url}
\begin{document}
%

\title{An Internet of Things Service Roadmap
%
}
%
%
%
%
%
\numberofauthors{1} 
%
\author{
%
%
\alignauthor
Athman Bouguettaya, Quan Z. Sheng, Boualem Benatallah, Azadeh Ghari Neiat, Sajib Mistry, Aditya Ghose, Surya Nepal, Lina Yao
}

\maketitle
\begin{abstract}
We propose a roadmap for leveraging the tremendous opportunities the \textit{Internet of Things (IoT)} has to offer. We argue that the combination of the recent advances in \textit{service computing} and \textit{IoT technology} provide a unique framework for innovations not yet envisaged, as well as the emergence of yet-to-be-developed IoT applications. This roadmap covers: emerging novel \textit{IoT services}, articulation of major research directions, and suggestion of a roadmap to guide the IoT and service computing community to address key IoT service challenges.



\end{abstract}
\section{Introduction}
The Internet of Things (IoT) is taking the world by storm, thanks to the \textit{proliferation of sensors} and \textit{actuators} embedded in everyday \textit{things}, coupled with the wide availability of high-speed Internet \cite{15} and evolution of the fifth generation (5G) networks \cite{37}. IoT devices are increasingly supplying information about the physical environment (e.g., infrastructure, assets, homes, and cars). The advent of IoT is enabling not only the \textit{connection} and \textit{integration} of devices that monitor physical world phenomena (e.g., temperature, pollution, energy consumption, human activities, and movement), but also data-driven and AI-augmented \textit{intelligence}. At all levels, synergies from advances in IoT, data analytics and AI are firmly recognized as strategic priorities for digital transformation \cite{6,12,15}.


IoT poses two key challenges \cite{4}: (1) \textit{communication} with things and (2) \textit{management} of things \cite{6}. The service paradigm is a key mechanism to overcome these challenges by transforming IoT devices into IoT services, where they will be treated as \textit{first-class} objects through the prism of \textit{services} \cite{13}. In a nutshell, services are at a higher level of abstraction than data. Services descriptions consist of two parts: functional and non-functional, i.e., Quality of Service (QoS) attributes \cite{22}. Services often transform data into an \textit{actionable knowledge} or achieve physical state changes in the operating context \cite{13}. As a result, the service paradigm is the perfect basis for understanding the transformation of data into actionable knowledge, i.e., making it useful. Despite the increasing uptake of IoT services, most organizations have not yet mastered the requisite knowledge, skills, or understanding to craft a successful IoT strategy. As a result, we do not have an adequate understanding of the ways by which we might leverage IoT opportunities.


From a service engineering perspective, IoT services may present difficult challenges, with many unsolved theoretical and technical questions \cite{13}. Such challenges stem from the scale of the systems contemplated, changes in service environments, quality of generated data and enabled services, the inherent heterogeneity and uncertainty of ubiquitous environments including connectivity, and growing concerns about the unintended consequences of the digital age: security and privacy breaches \cite{16}. For example, IoT devices can crowdsource a wide range of service types such as computing services, wireless energy sharing services, and environmental sensing services to other IoT devices in close proximity. In energy sharing services \cite{9035465}, IoT devices can wirelessly send energy to other nearby devices. However, because IoT services are crowdsourced, they are highly susceptible to improper and malicious usage. Stealing credit card information and sensitive medical histories, cyber attacks, denial of service attacks, and privacy violations are examples of improper and malicious usages of IoT services \cite{2}. 

We see the evolution of the work in IoT services as mirroring in a way, at least conceptually, the work done in the World Wide Web. These efforts over the last thirty years led to generic abstractions and computation techniques that enabled a holistic computing environment in which users, information, and applications establish on-demand interactions, to realize useful experiences and to obtain services. The benefit of such an environment originates from the \textit{added value} generated by the possible interactions. We believe that IoT services will require similar building blocks in terms of useful models and techniques to build the \textit{added-value promised by} the \textit{ubiquity} as well as the \textit{serendipity} of IoT services. We also believe that providing enhanced simplicity, agility, efficiency, and robustness in engineering and provisioning of IoT services will \textit{unlock the IoT service paradigm at a global scale}. The realization of this vision, however, poses formidable computing challenges to bring IoT services to the masses. While initial research outcomes exist which could be leveraged, significant progress is needed to make IoT services a tangible reality.

The remainder of the paper is organized as follows. In Section 2, we identify key criteria of IoT services via an analogy analysis of the Internet and the Web. In Section 3, we discuss the emerging technologies for the IoT environment. In Section 4, we describe the major challenges in IoT services and present a research roadmap for the identified challenges. In Section 5, we conclude and provide our assessments of the prospects for success.

\section{An Analogy Analysis}
We argue that for IoT to reach its full-potential (from ``\textit{technology}" to ``\textit{services}"), there is a need to analyze similar trajectories of other recent technological trends. We propose that there is such an analogy with the Internet and the Web (see Figure 1). While the Internet was created as a ``technology" for worldwide digital communication, the Web has \textit{transformed} the Internet into \textit{meaningful} services \cite{17}. We identify three key impacts of Web over the Internet:
\begin{itemize}
\item \textbf{Democratization}: The term ``\textit{democratization}" has its roots in political science and refers to the process of transitioning to a democratic form of government. More generally, it can be thought of as the \textit{process of removing the barriers} of privilege and of offering equal rights, access, and authority. The World Wide Web (WWW) achieved the democratization of access to information. Before the advent of the WWW, information often resided in repositories with privileged access or in places where the barriers to access were onerous. The Web ``democratized" access to information by removing (for the most part) barriers to access. In addition to democratizing access to information, the Web has also democratized the ability to publish information. Almost anyone can create a website and post information on it - those simple steps making that information accessible to everyone. In the early Internet, the information flow was in only one direction, which was static, with no way for users to add to or interact with the information. However, the Web emphasizes the importance of people's interactions with the Internet. Everyone has an opportunity to contribute to the Web. And, by paying attention to what users are looking for and doing online, better services (e.g., recommender systems) are designed over the years \cite{18}.

\item \textbf{Commoditization}: The Web \textit{redefines} the way businesses were performed. The Web enables e-Commerce platform technology using the Internet as a backbone and gives birth to today's platform economy. The platform technology has profoundly affected everyday life and how business and governments operate. Commercial transactions are conducted in electronic marketplaces that are supported by the platform technology. Transaction-oriented marketplaces include large e-malls, consumer-to-consumer auction platforms, multichannel retailers, and many millions of e-retailers. Examples of popular such transaction platform include Amazon, Airbnb, Uber, and Baidu. Massive business-to-business marketplaces have been created on the Web \cite{19}. Moreover, the platform economy enables more efficient use of resources. Almost \textit{instantaneous access} to services is made available by on-demand platforms using the Web. Such service orientation is not possible using only the concept of the Internet. Consumers have access to services and products from anywhere, and the price becomes a core factor in decision making. The bookstores are closing down; big retails are complaining about the penetration of online services such as Amazon; the emergence of Uber is disturbing the taxi industry. Established brick and mortar companies are competing to get a share of online customers. For example, big hotel chains are competing with Airbnb that does not own a single house. Every sector from IT to retails are affected by this phenomenon. 

\item \textbf{Digitization}: Digitization is \textit{the process of converting information} into digital formats. The Web is the single most important enabler for digitizing the information \cite{17}. The Web played a key role in moving the world from analog to digital form due to the increasing access to digital information. The advantages offered by digitization are the increasing access and \textit{preservation} of information. Moreover, digitization enables \textit{enhanced} services. Innovative services can be created using existing digital information in response to user demands. Such services have a direct impact on several industries. Libraries have been closing down; traditional print media companies are struggling to survive in the face of social media, and online blogs and forums; more people consume news from Twitter and Facebook than newspapers and television. Policies are debated and elections are fought on social media. A huge amount of information is being created and consumed in digital form directly. State libraries are collecting social media data for information preservation. Our daily life activities are stored and shared in real time. Every individual, organization, industry, and government is impacted by this transformation. 
\end{itemize}
\begin{figure}[t!]
\label{figanalogy}
\center
\includegraphics[width=8cm]{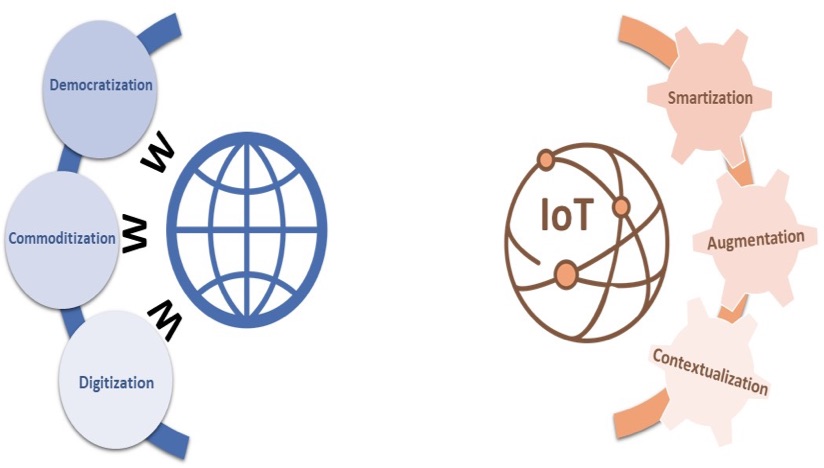}
\caption{Analogy between the Internet vs. the Web and IoT vs. IoT service}
\vspace{-4mm}
\end{figure}

There is a striking analogy of the IoT/services with the Internet/Web as services are the technology that transforms the IoT into a meaningful and useful framework. We outline three key criteria for defining novel IoT services, as shown in the right part of Figure 1.

\begin{itemize}

\item \textbf{Smartization}: Smartization refers to the \textit{process of introducing intelligence} to traditional systems to achieve sustainable, efficient, and convenient services. IoT services are instrumental in achieving this goal by working collaboratively to enable smart services for the people \cite{2}. Examples of such systems are smart cities, smart homes, and smart health. Consider an example of a smart city. The rapid industrialization has built global mega cities. These cities are now going through the post-industrialization development phase, where efficiency, sustainability, and livability become important factors for economic growth. These factors are largely addressed through smart city initiatives. For example, Virtual Singapore application enables city planners in Singapore to simulate various scenarios including emergency evacuation. Similarly, smart transportation systems will automate our roadways, railways, and airways, transform passenger experiences, and reshape the way cargo and merchandise are tracked and delivered. 

We believe the IoT service is the \textit{foundation} to build next-generation intelligent systems and to transform all aspects of our life. In the process of ``smartization", IoT services bring together data, analytics, and decision-making services within a single platform and ensure that they can work seamlessly and ubiquitously to provide an enhanced user experience.  Hence, the smartization plays a significant role in making intelligent cyber-physical systems. 

\item \textbf{Augmentation}: Augmentation refers to the \textit{process of creating new services} on demand by analyzing interactions among the devices and human to enhance the human experience. The self-driving vehicles, robots in aged care, and home automation are some examples of augmentation using IoT services. Augmentation has become a reality due to the availability of Artificial Intelligence (AI) assisted Intelligent Assistance (IA) \cite{5,14}. We believe that the future IA learns human behaviors, attitudes, and emotions, and creates new services on-demand to meet individual's needs.

One emerging application area that will be enhanced by IoT services is \textit{Augmented Reality (AR)}. IoT services enabled AR can be used to visualize and interact with data from thousands of sensors simultaneously in real-time. With such services, for example, a farmer can walk through his farms and get all information about soil, crops, water, moisture, temperature, and pest in real time with precise location. The farmer can interact with such services and get even better insights on demand in real-time.

We believe the IoT service is the key enabler to build intelligent systems on-demand \cite{20}. IoT services move the process of digitization from the artefacts in the Internet era to ``everything" in the IoT world - our cities, hospitals, transport systems as well as human beings. The IoT is a key enabling pillar of digitizing ``everything" since ``things" in the systems have \textit{embedded capability} to collect data, which captures the holistic view of the systems. The digitization process goes beyond the physical systems. For example, the ``brain wearables" help to digitize people's thoughts, feelings, and emotion and understand the neuroplasticity of our brain. These IoT services are very helpful to develop advance augmented tools for people suffering from many physical and mental sickness (e.g., anxiety, depression, paralysis). Hence, the \textit{augmentation} plays a significant role in enhancing interactions between human and physical systems. 

\item \textbf{Contextualization}: Contextualization refers to IoT services \textit{being aware of the situation} and quickly adapting to the environment. The adaptation is not only limited to transforming or filtering sensor data in a meaningful and useful way to fit for the purpose, but also instantiating appropriate actuators. Giving contexts to IoT services through sensor data and actuator actions becomes an important criterion to build \textit{personalized services} \cite{2,6,14}.

Data is the present whilst \textit{context} is the future! Context will be the key to all industries. For example, a service will be created on demand to help a customer to buy milk to fit with his dietary requirement and health conditions. Similarly, a personalized service will help people to pack their luggage while going on holidays - that is, one's suitcase, clothes and weather forecast can interact with each other as IoT services and create a new advisory service on demand. A personalized temperature service is created on-demand for an individual to maintain the room temperature at home depending on individual's preferences, meaning that an air-conditioning unit in the wall can interact with wearable sensors and create a personalized service.

A large number of start-up companies have emerged in recent times which are \textit{driven by contextualization}. This indicates that contextualization will grow exponentially in the coming years and most of our current products and services will be personalized. Hence, we believe that contextualization plays a significant role in enhancing the user experience by creating personalized services on demand in real-time \cite{6,14,20}.
 
\end{itemize}

The service computing research community has been continuing to design and develop IoT services for the last decade. Although there are incremental advancements, we argue that service computing has not fully explored to its potential in designing IoT services. This roadmap aims at outlining the vision and the underlying IoT service research challenges. 

\section{Emerging Technologies and IoT Services}
The advancements of existing computing paradigms such as data science, deep learning, and cloud computing and emerging technologies such as Edge computing, 5G networks, and blockchain are creating opportunities for innovative IoT services. These different paradigms or technologies have been explored in the context of IoT applications and platforms and are equally important for IoT services. They need to be coordinated to develop a distributed and dynamic IoT service \cite{23}. However, coordinating different computing paradigm with IoT services poses several research challenges. For example, the integration of data science with IoT needs to solve several research issues including the heterogeneity of IoT data formats, the real-time analytics, the data provenance, the dynamic data management and the IoT application orchestration \cite{23}. Several research directions are proposed to address these challenges from the data science perspective \cite{23}. Enhanced software abstraction of the IoT computation units such as MicroELement (MEL) and standard data integration protocols have the potential to resolve the IoT data heterogeneity issue \cite{24}. A MEL consists of microelements such as data, computing, and actuators to deploy integration and computing solutions. 

The \textit{IoT service development platform} should have the ability to efficiently store and analyze real-time large IoT data streams from different types of physical and social sensors. The Edge computing is the potential research platform where IoT data are stored and analyzed at the edge of the IoT network instead of cloud services \cite{25}. However, integrating Edge computing in the IoT service has several key research issues: a) programmability, b) naming, c) network and resource constraints management, d) QoS reliability, and e) security \cite{26}. Programmability refers to the development of service on heterogeneous edge nodes. Several novel approaches such as the development of computing streams are proposed to address the programmability in Edge computing \cite{27}. Naming refers to the standard way to discover and to communicate with a large amount of IoT services. Traditional naming approaches such as Domain Name Search (DNS) or Uniform Resource Identifier (URI) are not capable to serve the dynamic and large number of IoT services. Hence, novel naming approaches are required for dynamic IoT services. Another key challenge in integrating IoT service with Edge computing is to enabling large computing task with the resource-constrained edge nodes. The computation tasks are not preferred to migrate to the cloud as it may increase the network latency and hinder real-time decision makings. The research community is addressing this research issue by proposing different edge architectures and distributed task scheduling models \cite{28}. The application programmers have difficulties in ensuring the QoS of the IoT services due to diverse Edge infrastructures and fault events. Hence, the application QoS requirements and the underlying edge and fog infrastructures should be considered in building a QoS-aware IoT services \cite{28}.

The \textit{IoT service infrastructure} produces a large amount of sensor data that need to be analyzed to bring smartization in different applications such as smart homes and smart cities. The deep learning is a powerful analytic tool to extract new features and to bring intelligence in real-world applications \cite{29}. However, integrating deep learning into IoT services has several key research issues: a) learning from noisy sensor data, and b) enabling resource-constrained edge computing for deep learning algorithms. Data preprocessing is an important step for deep learning approaches. As IoT data is heterogeneous and generated from different sources, the accurate preprocessing or data curation is complex for real-time services. To the best our knowledge, existing approaches propose to use layered based learning frameworks where intermediate features are learned in edge servers and the final output layer is processed in the cloud \cite{30}. New learning acceleration engines are proposed for edge servers \cite{31}. These approaches are yet to adopt the full potential of deep learning for IoT services.

Security, privacy and data trust is another key research issue for integrating the data science into the IoT service. It is proposed to adapt the Blockchain technology to bring data provenance in the IoT service \cite{23}. The Blockchain technology has the ability to create a trusted, decentralized and autonomous system. Several Blockchain based IoT application framework is proposed in the existing literature \cite{32}. However, integrating Blockchain in an IoT service has several challenges such as resource limitations, interoperability of security protocols and the dynamic trust management \cite{33}. 

Existing research roadmaps on IoT services mainly focus on utilizing emerging technologies from the data science perspective \cite{23,25,30,32}. We focus on integrating emerging technologies from the service computing perspective.

\section{Challenges in IoT Service \\Research: A Roadmap}
In this section, we develop a deeper understanding of the key research challenges in the IoT service space by exploring \textit{novel classes of functionality} in future IoT innovations. 

\subsection{Actuation}
The IoT will achieve the \textit{democratization of actuation}, i.e., invoking Internet-addressable things to take state-altering actions. Actuation has not received much attention in the current discourse on IoT but is likely to become a major focus of attention in the near future. Accessible actuation entails that the ability to use IoT devices to take action can, in principle, be made available to all. The ability to operate IoT-enabled home devices remotely is already a well-recognized use case. Actuation over the Internet (we might refer to this as \textit{open actuation}) will have far-reaching and game-changing consequences that we have not yet started to fathom \cite{14,15}. We have seen a simpler version of this phenomenon in \textit{tele-operation}, but the impact of open actuation will be orders of magnitude greater. The tele-operation is typically \textit{point-to-point}, i.e., an operator invokes operations on a single device. The open actuation can be \textit{point-to-multipoint}, where a single operator invokes multiple actuators over the IoT. The tele-operation is typically \textit{pre-configured}, i.e., a tele-operation link is set up between an operator and a device by prior design (and often with investments in the physical infrastructure to make the tele-operation possible). The open actuation can be \textit{emergent}. An operator might identify devices on the fly whose operation would help to achieve the operator's goals. A bespoke infrastructure for tele-operation is not necessary.

A \textit{combination} of sensing and actuation gives us the ability to monitor and manage physical systems. Remote management of physical systems over the IoT can lead to the \textit{crowdsourced models of managing physical infrastructures}. For instance, citizen groups might volunteer to manage specific civic spaces, such as a park or a community hall. For a park, they might be able to monitor turf health through sensors, while using remotely operated actuators such as sprinklers to water the turf when required. Citizen groups could manage neighbourhood safety through similar means.

We have witnessed an exponential growth of autonomous systems in the last decade leading to the industrial revolution to realize the vision of Industry 4.0 \cite{39}. These autonomous systems are equipped with sensors and actuators, and support its self-operation. Self-driving car is a good example of such system. Autonomous vehicles are also in operations in many research and commercial activities \cite{40}. Examples include autonomous vehicles in mining, robots in healthcare, an underwater vehicle in climate study, etc. These autonomous systems are expected to interact with each other as well as their physical environments, building an autonomous Cyber Physical System (CPS) \cite{41}.
 
The fine-grained IoT-enabled device-level levers for sensing and actuation will make automation far more ubiquitous. The democratization of the management of physical infrastructures will also enable greater delegation and autonomy. The services of physical devices could be globally shared. 

\subsection{Servitization}
Servitization involves \textit{the wrapping of an existing product or system in a service-oriented model}. The IoT service will lead to a greater, and potentially ubiquitous servitization. The IoT service can transform existing devices into ones that offer value-added services. In this regard, IoT devices can harness service-oriented notions of \textit{publication, discovery, and composition}\cite{13,15,20}. For example, servitization can enable IoT devices to publish their functionalities and QoS guarantees in device registries which can be searched to discover new devices and their associated services. In the case of composition, servitized IoT devices can be composed using new service composition techniques to obtain desired functionalities that meet the QoS constraints. The servitization can also lead to new \textit{conceptions} of markets which regulate the usage of devices. For instance, servitized IoT devices may form a market for carbon credits that incentivizes the use of more carbon-friendly devices in more eco-friendly ways. 

Governments around the world are struggling to deal with legal and social policies arising from the tremendous growth in the use of IoT devices in citizens' daily life. Though many of the policies from Internet governance could be applicable to IoT devices, it requires a new thinking due to the complexity, scale and heterogeneity they bring. Servitization of IoT devices plays an important role to fill the gap and build policy, regulation and governance for them \cite{38}. 

\subsection{IoT Services Discovery}
Future Internet of Things is expected to be 50 to 100 times bigger than the current Internet, and the environments interacted by dynamic IoT services also evolve constantly \cite{3,7}. We identify a new set of challenges for IoT service discovery to enable the querying of billions of IoT resources to find the right service at the right time and location. We identify two different techniques that an IoT service discovery approach can adopt. The first technique is \textit{semantic annotations} for IoT service descriptions and their associated sensory data. For instance, the OpenIoT project\footnote{www.openiot.eu} exploits a semantic sensor network (SSN) ontology from W3C for the sensor discovery and dynamic integration. The Hydra project\footnote{www.hydramiddleware.eu} adopts OWL (i.e., an ontology language for Semantic Web) and SAWSDL (i.e., a semantic annotation of WSDL) to semantically annotate IoT services. A number of ontologies have been proposed to represent IoT resources and services including Ontology Web Language for Things (OWL-T) \cite{maamar2020owl}, IoT-Lite Ontology \cite{lite}, Comprehensive Ontology for IoT (COIoT) \cite{tayur2019comprehensive} and IoT-Stream \cite{elsaleh2020iot}. The Sensor Modeling Language SensorML which is a part of the OGC sensor Web enablement suite of standards, supports semantic descriptions of IoT services based on standardized XML tags. However, given the diversity and rapid IoT technological advances, it is challenging to reach an agreement on a single ontological standard for describing IoT services, and to maintain it. Regarding IoT semantic reasoning, similar approaches to those described in \cite{maarala2016semantic, chen2020modeling} may be used. 
The second technique uses the textual descriptions associated with IoT devices to locate the IoT services. Examples of IoT service discovery approaches based on the textual description are MAX \cite{8}, and Microsearch \cite{10}. A research challenge is the \textit{natural order ranking} of IoT contents.

The natural order ranking sorts contents by their intrinsic characteristics, rather than their relevance to a given query. In large data collections where a massive number of entities may be relevant to a query, natural order ranking mechanisms become crucial to deliver the most relevant results. PageRank is a well-known natural order ranking mechanism, which orders Web pages based on their importance via link analysis. Due to the size of IoT, another promising direction is to develop new natural order ranking mechanisms for the IoT contents to provide an effective and efficient IoT service discovery \cite{7}. It is important to define the natural order that is applicable across heterogeneous IoT contents and has scalability. One potential solution could rely on the quality of service (QoS) metrics of IoT services. Another possible solution is to construct a network of hidden links between IoT services and apply link analysis algorithms which are similar to PageRank. Discovering implicit relationships among IoT devices has been reported in recent research work \cite{11}. Considering the aforementioned techniques, the further work is to develop scalable approaches for the IoT service discovery. 

\subsection{Security, Privacy and Trust}
IoT services become \textit{key pillars} of automation and augmentation. Building trust in IoT services is the key to their success. Building trusted ecosystems among IoT services needs appropriate security, privacy and trust measures between IoT services which are enabled by sensors and actuators, and their interactions with human being \cite{2,16}. Like all other Internet-based services in the past, IoT-based services are also being developed and deployed without security consideration. IoT devices are inherently vulnerable to malicious cyber threats because of the following reasons: (1) they do not have well-defined perimeters, (2) they are highly dynamic and heterogeneous, (3) they are continuously changing because of mobility; and (4) they cannot be given the same protection that is received by enterprise services. In addition, due to ``\textit{billions}" of such IoT services, traditional human interaction driven security solutions do not scale for security analysts or IoT service end-users to carry out security activities. Those activities may include approving the granting of permissions to IoT devices and setting up access control policies and configurations. The IoT enabled augmented and automated decision-making systems will also ``encourage" malicious cyber threats due to the high value of such systems. Hence, coordinated efforts are required from the research community to address resulting concerns \cite{2}.

\textit{Is there such a thing as privacy in IoT services}? With the prevalence of smart phones, social media and people who tend to share so much information directly or indirectly, some researchers are starting to assert that there is no such thing as true privacy in the digital world. The impact of a data breach in an individual's life and regular targeted e-Commerce activities by the corporates have strengthened the view that privacy is more important than ever in the presence of IoT. In the beginning, the privacy concerns were limited to data, i.e., personal records, images, video, etc. With the adoption of smart phones, the privacy concern is moved from data to physical location as the location-based services are collecting an individual's location in real time. With the emergence of ``brain wearable" technology, one would be able to read people's mind and capture thoughts, feelings, hence raising the concern of mental privacy.

The technology trends in security are moving in two \textit{conflicting directions} in terms of IoT services. On the one hand, the advancement of quantum computing makes the current security technologies obsolete, as they can be broken within seconds. Consequently, we need to develop \textit{quantum resistance schemes}. On the other hand, current security technologies cannot be applied to many IoT systems, as they cannot operate on power constraints environment. As a result, the IoT services demand \textit{lightweight} quantum resistance security schemes. 

\subsection{Crowdsourcing IoT Services }
IoT devices are typically set up in fixed facilities or carried by humans. IoT users may crowdsource the functions of nearby IoT devices to suit their needs, such as WiFi hotspot sharing and wireless charging. The service paradigm can be applied as a key mechanism to abstract IoT devices and their functions along with their non-functional attributes (i.e., Quality of Service (QoS)) as crowdsourced IoT services from IoT users' perspectives. These services will run as proxies of IoT devices. Crowdsourcing IoT services is a new and promising direction for the IoT service platform \cite{13}. Since crowdsourcing is more likely to be used if there are financial rewards and other incentives, an appropriate incentive model is required to motivate IoT service providers to form various types of crowdsourced IoT service \cite{35}.

The interactions among crowdsourced IoT services has a greater complexity than traditional service-oriented applications due to a large number of the expected IoT applications in the crowdsourced environment. This induces some unique challenges on trust management for crowdsourced IoT services. Firstly, the expected large number of newly deployed IoT services will likely have historical records to show any initial trustworthiness credentials. Therefore, traditional feedback-driven trust management would not be a realistic approach for crowdsourced IoT environments. In this regard, trust management of crowdsourced IoT services requires an alternative trust anchor instead of historical records. It can be IoT services' inherent characteristics, which can generally reflect their trustworthiness. For example, IoT devices that are manufactured under a high-level security standard by a reputable manufacturer, are likely to be safely employed. The manufacturer's reputation can be the trust anchor of IoT devices. There may exist multiple trust anchors (e.g., the reputations of the manufacturers or owners of IoT devices), the aggregation of which would reflect the overall trustworthiness of IoT services. Secondly, the sheer diversity coupled with the expected large number of IoT application scenarios will redefine dynamism in service trustworthiness. The trustworthiness of an IoT service is greatly influenced by its application contexts and service users' trust preferences and requirements. As a result, the trust management in IoT environments should address the dynamic and fluid nature of IoT services. Thirdly, the traditional centralized trust management would be quite costly and inefficient for crowdsourced IoT services because of the expected large number of IoT devices. The distribution of trust-related information on IoT services is expected to be decentralized. A key challenge for IoT service consumers is, therefore, the trustworthy access to reliable trust information for the IoT trust evaluation in a decentralized way.

\subsection{Experiential IoT Services }
Experiential computing deals with digitally represented human experiences in everyday activities through every day ``things" that have embedded computing capabilities. IoT services enable the realization of the vision of experiential computing by creating an experiential environment through the mediation of four dimensions of human experiences (i.e., time, space, actors, and things). In this environment, users are able to explore and experience everyday events from multiple perspectives and revisit these events as many times as they wish to obtain the desired results \cite{1}. The computation paradigm in such environments moves from the current data analytics to experience analytics, where the computation will be performed on digitally represented user experiences. This brings a number of new research challenges:
\begin{itemize}
\item \textit{Can my autonomous vehicle give me the same experience that was felt by someone else}? 
\item \textit{How can one generate an experience from a massive amount of data collected from IoT devices}?
\item \textit{Can one transfer his/her experience from one environment (e.g., home) to another environment (e.g., office)?}
\end{itemize}

\subsection{Requirements-driven IoT Service Design }
The challenge of designing a device infrastructure, composed of both sensors and actuators, is difficult. Although the designing problem has some similarities with the problem of requirements-driven service composition, there are significant differences \cite{20}. In the service composition problem, a catalog of services is available a priori. In requirements-driven IoT service design, there are challenging questions that need to be addressed as follows:
\begin{itemize}
\item What are the \textit{data requirements} of the problem? What data would the decision modules and actuators need to be able to deliver on the required functionality? In the era of data analytics and the deployment of sophisticated AI systems, these are non-trivial problems, e.g., the challenges associated with \textit{feature engineering}. The many-to-many mapping between requirements and data items can be complex and requires equally complex reasoning to compute.

\item What collections of sensors will be necessary to meet the data requirements of the problem? What hardware configurations would support the relevant non-functional requirements? Where should sensors be located? What hardware performance guarantees would be necessary to ensure that overall system-level performance guarantees are met?

\item In a similar spirit, what actuators would the system require? What locations would be appropriate, what hardware configurations would be necessary and what hardware performance guarantees would satisfy the overall non-functional requirements?

\item What kinds of coordination models would be necessary to orchestrate the behaviors of sensors and actuators to meet the stated requirements? Will existing schemes for specifying coordination models (such as process modeling notations) suffice?
\end{itemize}

\subsection{Computing Complex Compositions of Sensors and Actuators}
As discussed above, the problem of the IoT system design takes us into uncharted territory. The hardware dimensions of the problem, i.e., finding the appropriate hardware configurations for sensors and actuators and the spatio-temporal dimensions need to be integrated and addressed \cite{20,34}. Furthermore, the Internet-of-Everything (IoE) aspects \cite{36} add greater complexity. The autonomous human elements of the system and the AI-enabled computation components whose behavior would be emergent and not entirely predictable at design time need to be considered. In this regard, we identify the following challenges: 
\begin{itemize}
\item \textbf{Managing resource-constrained sensing and actuation}: IoT systems often need to operate under significant resource constraints. This necessitates a significant re-working of standard approaches to system design, which leads to a novel conception of \textit{resource-aware design}. In the spirit of earlier thinking on sensor networks, we need to design sensing behaviors that take into account finite energy reserves on sensor batteries. Similarly, actuator behavior would need to account for the finite capacity of actuator power sources. 

\item \textbf{Managing sensors and actuators at scale}: The IoT will enable us to address individually (e.g., resource locators) devices at a very fine-grained level, and consequently on a very large scale \cite{7,12,21}. However, system design and management might not be very effective at these levels of granularity. In some cases, assigning individual addresses or resource locators at these low levels of granularity might also be challenging. We will, therefore, require novel abstractions that allow us to aggregate (and dis-aggregate) groups of sensors and actuators. An example is abstractions for classes of mutually interchangeable sensors and actuators. Interchangeability could be parametric. A set of sensors could be swapped for each other under a given set of functional requirements but not for a different set of functional requirements. Protocols for invoking sensor or actuator behavior will also need to exploit these abstractions. A range of similar issues also needs to be addressed for managing IoT devices at scale.

\end{itemize}

\subsection{Large-Scale IoT Experimental Facilities}
While IoT-based digital strategies and innovations provide industries across the spectrum with exciting capabilities to create a competitive edge and build more value into their services, as what the Internet has done in the past 25 years, there are still significant gaps in making IoT a reality. One such gap lies on the missing of a large-scale, real-world experimental testbed for research and experimentation of new IoT service technologies \cite{6,12}. 

The current IoT research infrastructures are largely in small scale, fragmented. There are not, therefore, suitable for IoT research and development. There is an urgent need to create such a unique research facility to stimulate advanced experimental research and realistic assessment of IoT technologies. Fueling the use of such a facility among the scientific community, end users, and service providers would increase the understanding of the technical and societal barriers in IoT adoption. The IoT-Lab\footnote{https://www.iot-lab.info} is a recent effort in this trend. IoT-Lab test beds are located at six different sites across France, which are publicly accessible. A similar effort is also currently happening in Australia, aiming at establishing a nation-wide IoT testbed across seven sites in major Australian cities. Digital Twins is a recent technology  development that have attracted both industry and academia, and can be exploited to build large-scale IoT experimental facilities\footnote{https://azure.microsoft.com/en-au/services/digital-twins/
}.

\subsection{IoT Data Analytical Services}
The IoT analytics aims at delivering domain-specific solutions by aggregating and distilling heterogeneous IoT data to obtain information and actionable knowledge of appropriate quality and integrity. There is a need for a new paradigm of advanced IoT analytical services, which effectively and efficiently provide the underlying intelligence via harnessing the combination of physical and cyber worlds to turn IoT data into IoT intelligence. The following are some of the key identified challenges:
\begin{itemize}
\item \textbf{Dynamic contextual changes}: IoT data are tightly associated with multi-faceted dynamic contexts, including user's internal contexts (e.g., users' activities), external contexts (e.g., location and time), and things' contexts (e.g., expiration, usage status, and locations). Therefore, the effective IoT data analytical services are required to be capable of capturing both the salient changes and subtle ones of real-time contexts. 

\item \textbf{Tangled complex relationships}: IoT data exhibits highly heterogeneous and multi-dimensional correlations. For instance, user behaviors on things are intrinsically correlated both spatially and temporally \cite{11}. The new paradigm of IoT data analytical services needs to decode and leverage the heterogeneous nature of complex relationships. 

\item \textbf{Real-time distributed analytics}: IoT data are generated with high volume from scattered sources on a continuous basis, and the value of data might exponentially decay over timestamps for many IoT applications. This requires the analytical models to derive useful patterns and actionable knowledge with quality summarizations and then use these for provisioning streaming IoT analytics.  

\item \textbf{Reducing bias and ensuring fairness in IoT data analytics}: IoT data analytics would heavily rely on advanced machine learning techniques. The fairness in AI/ML technologies is an active research in itself and several techniques have been developed \cite{bellamy2019ai}. However, we need to understand what it means to IoT services, and should apply the same rigorous scrutiny to IoT data and services.

\end{itemize}

\section{conclusion}
The IoT is widely considered as a new revolution of the Internet where billions of everyday objects are connected to empower human interactions with both virtual and physical worlds unprecedentedly. We believe that advancements made in the service computing over the past decades have not fully explored its potential in the designing of IoT services. We have identified three key criteria that define IoT services, namely smartization, augmentation, and contextualization. We outlined ten main challenges in developing an IoT service. Designing and engineering scalable and robust IoT based solutions remains a deeply challenging problem.  We identify critical directions spanning discovery, security, privacy and analytics. Interesting future research directions include: 
\begin{itemize}

\item Actuation over the Internet should be further investigated to provide ubiquitous automation.
\item Existing policies and regulations for Internet governance should be enhanced to enable servitization of IoT devices.
\item IoT service discovery approaches should be dynamic and scalable to cater to the gigantic size and diversity of IoT and rapid IoT technological advances.
\item Lightweight quantum security schemes should be explored for power-constrained IoT services.
\item The trust management framework for crowdsourcing IoT services should be decentralized to manage the dynamism of service trust.
\item Data analytics approaches should be translated to experience analytics to create an experiential IoT environment.
\item Complex feature engineering should be investigated to address requirements-driven IoT service design.
\item Computing complex compositions of sensors and actuators should be AI-driven and follow the resource-aware design.


\end{itemize}

\bibliographystyle{abbrv} 
\bibliography{bibs4cacm}  
%
\balancecolumns
\end{document}